\documentclass[twocolumn,showpacs,preprintnumbers,amsmath,amssymb,prb]{revtex4}

\usepackage[dvips]{graphicx}
\usepackage{dcolumn}
\usepackage{bm}

\begin{document}

\title{Quasiclassical theory of superconducting states under magnetic fields:\\
Thermodynamic properties}
\author{Hiroaki Kusunose}
\affiliation{Department of Physics, Tohoku University, Sendai 980-8578, Japan}

\date{January 7, 2004}

\begin{abstract}
We present a simple calculational scheme for superconducting properties under magnetic fields.
A combination of an approximate analytic solution with a free energy functional in the quasiclassical theory provides a wide use formalism for spatial-averaged thermodynamic properties, and requires a little numerical computation.
The theory covers multiband superconductors with various set of singlet and unitary triplet pairings in the presence of an impurity scattering.
It is also applicable to analyze experimental results in a rotating magnetic field with help of band structure calculations.
We demonstrate the application to $s$-wave, $d_{x^2-y^2}$-wave and two-band $s$-wave pairings, and discuss the validity of the theory comparing with previous numerical studies.
\end{abstract}
\pacs{74.20.-z,74.25.-q,74.25.Bt}

\maketitle

\section{introduction}
A magnetic field has been a good probe to investigate a gap symmetry of superconductivity.
In particular, the low-energy excitations in the mixed state exhibit unconventional behaviors inherent in different nature of a vortex core, nodal structure and quasiparticle transfer between vortices \cite{Vekhter01,Joynt02,Mackenzie03,Choi02}.
In a recent development in multiband superconductors \cite{Mackenzie03,Choi02,Ohashi02,Koshelev03,Miranovic03a,Dahm03,Zhitomirsky04}, magnetic fields play an important role, e.g., weak magnetic fields suppress strongly smaller gap of the multiband superconductivity leading to drastic enhancement in field dependence of the residual density of states \cite{Bouquet02,Nakai02}, the thermal conductivity \cite{Sologubenko02,Kusunose02} and so on.
Moreover oriented field measurements have performed in order to determine directly superconducting gap structure \cite{Izawa02a,Izawa02b,Izawa01a,Izawa01b,Deguchi03,Aoki03,Park03a,Park03b,Park03c,Deguchi04}.

From the theoretical side various approaches have been made to investigate bulk properties under magnetic fields.
The Ginzburg-Landau theory is powerful but is restricted to regions near the critical fields.
The Volovik theory is used frequently, in which the quasiparticle excitation energy is approximated by taking into account the Doppler shift of the local superconducting current \cite{Vekhter01,Volovik93,Franz99,Vekhter99a,Won01}.
This method neglects the scattering by vortex cores and the overlap of the core states, which makes the theory to be valid only at low-temperature and low-field regions \cite{Dahm02}.
Meanwhile, fully numerical calculations for the quasiclassical transport-like equation or the Bogoliubov-de Gennes equation have been performed \cite{Klein87,Pottinger93,Ichioka97,Ichioka99,Ichioka02}.
Although these approaches are necessary to discuss local structure of the core states, it is not easily applicable for transport properties in the linear-response theory, impurity effect and extension to multiband superconductors with detailed band structure because of intractable aspect of numerical computation.

Motivated by these situation, it is useful to arrange a semi-analytic approach to analyze various experimental results and to check their consistency on an equal footing.
For weak-coupling superconductors, there exists a reliable approximate analytic solution \cite{Brandt67,Pesch75} in the quasiclassical formalism \cite{Eilenberger68,Larkin68} with a certain kind of mean-field approach.
Although required conditions seem to be valid near the upper critical field $H_{c2}$, it is meaningful to extrapolate the theory to lower fields.
With understanding of its validity, it becomes very useful to discuss spatial-averaged thermodynamic properties of superconductivity under magnetic fields.
The theory covers multiband superconductors with various set of singlet and unitary triplet pairings in the presence of an impurity scattering.
The realistic band structure can be included in arbitrary direction of magnetic fields.
Moreover it is also applicable to investigate linear-response coefficients in a similar manner \cite{Klimesch78,Serene83,Rammer86,Houghton98,Vekhter99}, which will be presented in a subsequent paper.

The paper is organized as follows.
In the next section, we introduce the quasiclassical equations in a general form, then we give explicit expressions of the transport-like equation and the free energy functional for singlet and unitary triplet pairings.
In \S3 we explain the approximate analytic solution for quasiclassical equations in the presence of the impurity scattering.
Then we express useful formulas for basic thermodynamic quantities.
In \S4 we extend the theory to multiband superconductors and present a set of formulas in terms of dimensionless parameters for the convenience of the users.
The application of the method to $s$-wave, $d_{x^2-y^2}$ and two-band $s$-wave pairings and the comparison with previous numerical solutions are given in \S5.
The last section summarizes the paper.

\section{quasiclassical equations}
The microscopic Green's function contains all the information about the single-particle properties. In most cases, however, states far from the Fermi energy or rapid oscillations inherent in the Fermi surface play little role to superconducting properties.
The key idea of the quasiclassical theory is that those irrelevant states are integrated out from the beginning in the Gor'kov formalism \cite{Eilenberger68,Larkin68,Serene83,Rammer86}.
Then, the basic equations of the quasiclassical theory give a complete description of the superconducting properties on a coarse-grained scale.
In this section, we first review the derivation of the standard quasiclassical equations under magnetic fields, and we introduce necessary notations through this step.
Next, we restrict ourselves to the cases of the singlet and the unitary triplet pairings, and then we construct a free energy functional as a generating functional for the quasiclassical formalism.

\subsection{Derivation of quasiclassical equations}
Let us start with the mean-field (MF) Hamiltonian in the $4\times4$ Nambu-Gor'kov space with a spatial inhomogeneity ($c=\hbar=k_{\rm B}=1$ hereafter),
\begin{equation}
H_{\rm MF}=\frac{1}{2}\int d{\bm x}_1 d{\bm x}_2
{\bm \Psi}^\dagger({\bm x}_{1})\biggl(
\hat{K}^{12} + \hat{\Delta}^{12}
\biggr)\hat{\rho}^z {\bm \Psi}({\bm x}_{2}),
\label{mf-ham}
\end{equation}
where the creation-annihilation field operators at the position ${\bm x}_s$ are composed as
${\bm \Psi}^\dagger({\bm x}_{s})=[\psi^\dagger_\uparrow({\bm x}_s),\psi^\dagger_\downarrow({\bm x}_s),\psi_\uparrow({\bm x}_s),\psi_\downarrow({\bm x}_s)]$.
An operator with a hat acts on the Nambu-Gor'kov space, and $\hat{\bm\rho}$ ($\hat{\bm\sigma}$) is the vector composed of Pauli matrices on the $2\times2$ particle-hole (spin) space remaining the other degrees of freedom unchanged.
The unit matrix is denoted by $\hat{1}$.
The kinetic-energy matrix is then given by
\begin{equation}
\hat{K}^{12}=\delta({\bm x}_1-{\bm x}_2)\xi\left(
-i{\bm\nabla}_2\hat{\rho}^z-e{\bm A}_2\hat{1}
\right),
\end{equation}
with a single-particle energy operator $\xi({\bm p})$, where ${\bm A}_s$ is the vector potential at ${\bm x}_s$ and $e<0$ is the electron charge.
The superconducting gap matrix is given by
\begin{equation}
\hat{\Delta}^{12}=\left[
\begin{array}{cc}
0 & \Delta({\bm x}_1,{\bm x}_2) \\
-\Delta^\dagger({\bm x}_2,{\bm x}_1) & 0
\end{array}\right],
\end{equation}
and its component is defined in terms of the two-body interaction $V_{\alpha\beta\alpha'\beta'}({\bm x}_1,{\bm x}_2)$ as
\begin{equation}
\Delta_{\alpha\beta}({\bm x}_1,{\bm x}_2)=-\sum_{\alpha'\beta'}V_{\alpha\beta\alpha'\beta'}({\bm x}_1,{\bm x}_2)\langle \psi_{\beta'}({\bm x}_2)\psi_{\alpha'}({\bm x}_1)\rangle.
\end{equation}
Corresponding to the MF Hamiltonian, the Nambu-Gor'kov equation reads
\begin{multline}
\int dy \left[-\delta(x_1-y)\hat{\rho}^z\frac{\partial}{\partial\tau_1}-
\delta(\tau_1-\tau)\left(\hat{K}^{1y}+\hat{\Delta}^{1y}\right)
\right.\\ \left.
-\hat{\Sigma}^{1y}(\tau_1-\tau)\hat{\rho}^z\right]
\hat{\rho}^z\hat{G}({\bm y},{\bm x}_2;\tau_1-\tau_2)=\hat{1}\delta(x_1-x_2),
\label{dyson-space}
\end{multline}
for the imaginary-time Green's function
\begin{equation}
\hat{G}({\bm x}_1,{\bm x}_2;\tau_1-\tau_2)=-\langle T_\tau \Psi({\bm x}_1,\tau_1) \Psi^\dagger({\bm x}_2,\tau_2)\rangle,
\label{greens-func}
\end{equation}
where we have used $x_s=({\bm x}_s,\tau_s)$ and $y=({\bm y},\tau)$.
Here the operator $T_\tau$ arranges the field operators in ascending order of the imaginary time $0<\tau<1/T$.
The self-energy $\hat{\Sigma}$ has been taken into account to treat the impurity scattering in subsequent discussion.

In order to integrate out the rapid oscillations, we introduce the center-of-mass coordinate, ${\bm R}=({\bm x}_1+{\bm x}_2)/2$ and the relative coordinate, ${\bm r}={\bm x}_1-{\bm x}_2$, and perform the Fourier transformation in the latter according to
\begin{equation}
f_{\bm R}({\bm k};i\omega_n)=\int d{\bm r}\int_0^{1/T} d{\tau} f\left({\bm x}_1,{\bm x}_2;\tau\right)
e^{-i({\bm k}\cdot{\bm r}-\omega_n\tau)},
\end{equation}
where $\omega_n=(2n+1)\pi T$ is the fermionic Matsubara frequency.
A product of functions, $h^{12}(\tau_1-\tau_2)=\int d{\bm y}\int d\tau f^{1y}(\tau_1-\tau)g^{y2}(\tau-\tau_2)$, is then transformed into so-called ``circle-product" \cite{Eckern81},
\begin{multline}
h_{\bm R}({\bm k};i\omega_n)=
\exp\left[\frac{1}{2i}({\bm\nabla}_{{\bm k}'}\cdot{\bm\nabla}_{\bm R}-{\bm\nabla}_{\bm k}\cdot{\bm\nabla}_{{\bm R}'})\right]
\\ \times
f_{\bm R}({\bm k};i\omega_n)g_{{\bm R}'}({\bm k}';i\omega_n)\biggr|_{
{\bm k}'={\bm k}, {\bm R}'={\bm R}
}.
\end{multline}
In translationally invariant systems, the internal magnetic field, ${\bm B}={\bm\nabla}\times{\bm A}$ is only the source of slowly varying field dependence.
In the lowest order in ${\bm\nabla}_{\bm R}$, which is necessary in this paper, the circle-product is given simply by the product of each Fourier transformations, $h_{\bm R}({\bm k})\sim f_{\bm R}({\bm k})g_{\bm R}({\bm k})$.
Noting the transformation,
\begin{multline}
\xi(-i{\bm\nabla}_s\hat{\rho}^z -e{\bm A}_s)\to \\
\xi_{\bm k}+\frac{i}{2}{\bm v}(\hat{\bm k})\cdot\left[(-1)^{s}{\bm\nabla}_{\bm R} + 2ie{\bm A}_{\bm R}\hat{\rho}^z\right],
\end{multline}
in the leading order, we obtain the Nambu-Gor'kov equation as
\begin{multline}
\left[\left(
i\omega_n+e{\bm v}(\hat{\bm k})\cdot{\bm A}_{\bm R}
\right)\hat{\rho}^z
-\left(\xi_{\bm k}-\frac{i}{2}{\bm v}(\hat{\bm k})\cdot{\bm\nabla}_{\bm R}\right)\hat{1}\right.
\\
\left.
-\hat{\Delta}_{\bm R}(\hat{\bm k})-\hat{\sigma}_{\bm R}(\hat{\bm k};i\omega_n)
\right]\hat{\rho}^z\hat{G}_{\bm R}({\bm k};i\omega_n)=\hat{1},
\label{dyson-k}
\end{multline}
where $\hat{\bm k}={\bm k}_{\rm F}/|{\bm k}_{\rm F}|$ is the unit wave vector at the Fermi surface and ${\bm v}(\hat{\bm k})={\bm\nabla}_{\bm k} \xi({\bm k}_{\rm F})$ is the Fermi velocity.
Here we have considered that the self-energy in question has weak momentum dependence with respect to $k_{\rm F}$, hence we have denoted as $\hat{\sigma}_{\bm R}(\hat{\bm k};i\omega_n)=\hat{\Sigma}({\bm k};i\omega_n)|_{{\bm k}={\bm k}_{\rm F}}\hat{\rho}^z$.

To obtain a closed quasiclassical theory, we must construct equations for a quasiclassical propagator, $\hat{g}\sim \int d\xi \hat{G}$.
However, a simple integration of eq.~(\ref{dyson-k}) results in unphysical divergences due to $\xi_{\hat{\bm k}}\hat{G}$ term in the left-hand side and $\hat{1}$ term in the right-hand side.
We get rid of this inconvenience with the help of the ``left-right subtraction trick", which transforms the Dyson's equation into transport-like equations for $\hat{g}$.
The trick is to eliminate those divergent terms by subtracting the right-hand Dyson equation, in which operators in the brace of eq.~(\ref{dyson-space}) act on the second argument of $\hat{G}$ from the right.
A similar argument leading to eq.~(\ref{dyson-k}) gives the right-hand equation,
\begin{multline}
\hat{\rho}^z\hat{G}_{\bm R}({\bm k};i\omega_n)
\left[\left(
i\omega_n+e{\bm v}(\hat{\bm k})\cdot{\bm A}_{\bm R}
\right)\hat{\rho}^z
\right. \\ \left.
-\left(\xi_{\bm k}+\frac{i}{2}{\bm v}(\hat{\bm k})\cdot{\bm\nabla}_{\bm R}\right)\hat{1}
-\hat{\Delta}_{\bm R}(\hat{\bm k})-\hat{\sigma}_{\bm R}(\hat{\bm k};i\omega_n)
\right]=\hat{1}.
\end{multline}
The unphysical terms then cancel, and the terms left can be $\xi$-integrated to yield the following transport-like equations,
\begin{multline}
\left[
\left(
i\omega_n+e{\bm v}(\hat{\bm k})\cdot{\bm A}_{\bm R}
\right)\hat{\rho}^z-\hat{\Delta}_{\bm R}(\hat{\bm k})-\hat{\sigma}_{\bm R}(\hat{\bm k};i\omega_n)\;,\right.\\ \left.
\hat{g}_{\bm R}(\hat{\bm k};i\omega_n)
\right]
+i{\bm v}(\hat{\bm k})\cdot{\bm \nabla}_{\bm R} \hat{g}_{\bm R}(\hat{\bm k};i\omega_n) = 0,
\label{qc-eq}
\end{multline}
where we have defined the quasiclassical propagators as
\begin{equation}
\hat{g}_{\bm R}(\hat{\bm k};i\omega_n)=\int \frac{d\xi}{\pi} \hat{\rho}^z\hat{G}_{\bm R}({\bm k};i\omega_n).
\label{qc-propagator}
\end{equation}
At the subtraction step, a normalization of $\hat{g}$ is lost.
The appropriate normalization is known as $\hat{g}_{\bm R}^2(\hat{\bm k};i\omega_n)=-\hat{1}$, which can be confirmed explicitly in a homogeneous case without the internal field ${\bm B}_{\bm R}$ \cite{Eckern81}.

To complete the quasiclassical formalism, we give self-consistent equations for the gap and the superconducting current as,
\begin{multline}
\Delta_{{\bm R}\alpha\beta}(\hat{\bm k})=
-\pi N_0 T\sum_n\sum_{\alpha'\beta'}
\\ \times
\left\langle V_{\alpha\beta\alpha'\beta'}(\hat{\bm k},\hat{\bm k}')\hat{g}^{12}_{{\bm R}\alpha'\beta'}(\hat{\bm k}';i\omega_n)\right\rangle_{\hat{\bm k}'},
\label{gap-eq}
\end{multline}
\begin{multline}
{\bm\nabla}_{\bm R}\times \left( {\bm B}_{\bm R}-{\bm H}_{\bm R} \right) =
\\
4e\pi^2 N_0 T\sum_n\sum_\alpha
\left\langle {\bm v}(\hat{\bm k})\hat{g}^{11}_{{\bm R}\alpha\alpha}(\hat{\bm k};i\omega_n)\right\rangle,
\label{ampere-eq}
\end{multline}
where ${\bm H}_{\bm R}$ is the external magnetic field and $\hat{g}^{ij}$ is understood as the $(i,j)$-element in the particle-hole space.
The bracket $\langle\cdots\rangle=\int d\hat{\bm k}v^{-1}(\hat{\bm k})\cdots/\int d\hat{\bm k}v^{-1}(\hat{\bm k})$ represents the angular average over the Fermi surface and $N_0$ is the density of states (DOS) per spin at the Fermi energy.

\subsection{Cases of singlet and unitary triplet pairings and generating functional of quasiclassical theory}
In this subsection, we give only expressions for a unitary triplet pairing.
For a singlet pairing, vector quantities should be replaced by scalar ones and a unit matrix is used instead of ${\bm\sigma}$.

In the case of a unitary triplet pairing, the spin structure of the gap is decomposed as
\begin{equation}
\Delta_{{\bm R}\alpha\beta}(\hat{\bm k})={\bm \Delta}_{\bm R}(\hat{\bm k})\cdot({\bm\sigma}i\sigma^y)_{\alpha\beta},
\end{equation}
with $i({\bm \Delta}\times{\bm \Delta}^*)=0$ \cite{Sigrist91}.
The quasiclassical propagator also has the form,
\begin{equation}
\hat{g}_{\bm R}=\left[
\begin{array}{cc}
-ig_{\bm R}\delta_{\alpha\beta} & {\bm f}_{\bm R}\cdot({\bm\sigma} i\sigma^y)_{\alpha\beta} \\
-{\bm f}_{\bm R}^\dagger\cdot({\bm\sigma}i\sigma^y)^\dagger_{\alpha\beta} & ig_{\bm R}\delta_{\alpha\beta}
\end{array}\right].
\label{mat-g}
\end{equation}
By definitions (\ref{greens-func}) and (\ref{qc-propagator}), the normal and the anomalous propagators have the symmetry \cite{Serene83},
\begin{equation}
g_{\bm R}(\hat{\bm k};i\omega_n)=
-g_{\bm R}(-\hat{\bm k};-i\omega_n)=
g_{\bm R}^*(-\hat{\bm k};i\omega_n),
\label{sym-g}
\end{equation}
and
\begin{equation}
{\bm f}_{\bm R}(\hat{\bm k};i\omega_n)=
\mp{\bm f}_{\bm R}(-\hat{\bm k};-i\omega_n)=
\mp{\bm f}^{\dagger *}_{\bm R}(-\hat{\bm k};i\omega_n),
\label{sym-f}
\end{equation}
where the upper (lower) sign is applied for the triplet (singlet) state.
Note that the gap function satisfies ${\bm\Delta}_{\bm R}(\hat{\bm k})=\mp {\bm\Delta}_{\bm R}(-\hat{\bm k})$ similar to (\ref{sym-f}).
According to these relations, we consider only $\omega_n>0$ hereafter without loss of generality.

At this point we explicitly take into account an effect of impurity scattering.
Since an impurity potential is short range as much shorter than length scale of ${\bm R}$ dependence, we merely follow the standard $t$-matrix theory to treat non-magnetic impurities \cite{Shiba68,Schmitt-Rink86,Hirschfeld86}.
Decomposing the impurity self-energy similar to eq. (\ref{mat-g}),
\begin{equation}
\hat{\sigma}_{\bm R}=
\left[
\begin{array}{cc}
-i\sigma^{(n)}_{\bm R}\delta_{\alpha\beta} & {\bm \sigma}^{(a)}_{\bm R}\cdot({\bm\sigma} i\sigma^y)_{\alpha\beta} \\
-{\bm \sigma}^{(a)\dagger}_{\bm R}\cdot({\bm\sigma}i\sigma^y)^\dagger_{\alpha\beta} & i\sigma^{(n)}_{\bm R}\delta_{\alpha\beta}
\end{array}\right],
\end{equation}
we obtain
\begin{eqnarray}
&&
\sigma^{(n)}_{\bm R}(i\omega_n)=\frac{1}{2\tau_0}\frac{\langle g_{\bm R}\rangle}{\cos^2\delta+D_{\bm R}\sin^2\delta},
\\&&
{\bm\sigma}^{(a)}_{\bm R}(i\omega_n)=\frac{1}{2\tau_0}\frac{\langle {\bm f}_{\bm R}\rangle}{\cos^2\delta+D_{\bm R}\sin^2\delta},
\end{eqnarray}
with $D_{\bm R}=\langle g_{\bm R}\rangle^2+\langle {\bm f}_{\bm R}\rangle\cdot\langle {\bm f}_{\bm R}^\dagger\rangle$, where $\tau_0$ is the quasiparticle lifetime in the normal state and $\delta$ is the impurity phase shift (e.g. $\delta=0$ corresponds to the Born approximation and $\delta=\pi/2$ is the unitarity limit).
Here $\sigma^{(n)}_{\bm R}$ and ${\bm\sigma}^{(a)}_{\bm R}$ satisfy the symmetry relation similar to (\ref{sym-g}) and (\ref{sym-f}), respectively.
For convenience, we introduce the renormalized frequency and gap function as
\begin{eqnarray}
&&
\tilde{\omega}_{n{\bm R}}(i\omega_n)=\omega_n+\sigma^{(n)}_{\bm R},
\\&&
\tilde{\bm \Delta}_{\bm R}(\hat{\bm k};i\omega_n)={\bm \Delta}_{\bm R}(\hat{\bm k})+{\bm\sigma}^{(a)}_{\bm R},
\end{eqnarray}
then the self-energies can be absorbed formally into the ``tilde" quantities in quasiclassical equations.

The explicit expression of eq.~(\ref{qc-eq}) is then given by
\begin{eqnarray}
&&\tilde{\cal L}_+{\bm f}_{\bm R}=g_{\bm R}\tilde{\bm \Delta}_{\bm R}(\hat{\bm k};i\omega_n),
\;\;\;\;\;
\tilde{\cal L}_-{\bm f}_{\bm R}^\dagger=g_{\bm R}\tilde{\bm \Delta}^\dagger_{\bm R}(\hat{\bm k};i\omega_n),\nonumber \\
&&{\bm v}(\hat{\bm k})\cdot{\bm\nabla}_{\bm R}g_{\bm R}=\tilde{\bm\Delta}^\dagger_{\bm R}(\hat{\bm k};i\omega_n)\cdot{\bm f}_{\bm R}-{\bm f}_{\bm R}^\dagger\cdot\tilde{\bm \Delta}_{\bm R}(\hat{\bm k};i\omega_n),
\nonumber\\&&
\label{qc-eq2}
\end{eqnarray}
with the normalization, $g_{\bm R}^2+{\bm f}_{\bm R}\cdot{\bm f}_{\bm R}^\dagger=1$, where we have introduced
\begin{equation}
{\cal L}_\pm(\hat{\bm k},{\bm R};i\omega_n)=\omega_{n}\pm\frac{1}{2}{\bm v}(\hat{\bm k})\cdot\left[{\bm\nabla}_{\bm R}\mp2ie{\bm A}_{\bm R}\right],
\label{l-op}
\end{equation}
and $\tilde{\cal L}_\pm={\cal L}_\pm(\omega_n\to\tilde{\omega}_{n{\bm R}})$.
Note that one of three equations in (\ref{qc-eq2}) is independent due to the symmetry relations (\ref{sym-g}) and (\ref{sym-f}) and the normalization condition.
In the case of ${\bm B}_{\bm R}=0$, the solutions of eq.~(\ref{qc-eq2}) are given by
\begin{eqnarray}
&&
g=\left[1+\frac{\tilde{\bm\Delta}(\hat{\bm k};i\omega_n)\cdot\tilde{\bm\Delta}^\dagger(\hat{\bm k};i\omega_n)}{\tilde{\omega}_n^2}\right]^{-1/2},
\label{propa-g0}
\\&&
{\bm f}=\frac{\tilde{\bm\Delta}(\hat{\bm k};i\omega_n)}{\tilde{\omega}_n}g,\;\;\;\;\;
{\bm f}^\dagger=\frac{\tilde{\bm\Delta}^\dagger(\hat{\bm k};i\omega_n)}{\tilde{\omega}_n}g.
\label{propa-f0}
\end{eqnarray}

The interaction leading to the triplet (singlet) pairing has the form
\begin{equation}
V_{\alpha\beta\alpha'\beta'}(\hat{\bm k},\hat{\bm k}')=P^\pm_{\alpha\beta\alpha'\beta'}V(\hat{\bm k},\hat{\bm k}'),
\end{equation}
where $P^\pm_{\alpha\beta\alpha'\beta'}=(\delta_{\alpha\alpha'}\delta_{\beta\beta'}\pm\delta_{\alpha\beta'}\delta_{\beta\alpha'})/2$ is the projection operator to the triplet (singlet) state.
The interaction for the triplet (singlet) pairing is odd (even) against $\hat{\bm k}\to-\hat{\bm k}$ or $\hat{\bm k}'\to-\hat{\bm k}'$.
Then we classify the interaction and the gap function by irreducible representations of the point-group \cite{Sigrist91}.
Considering a certain irreducible representation $\Gamma$ with the highest transition temperature, $T_c$, we can write $V(\hat{\bm k},\hat{\bm k}')$ in a separable form,
\begin{equation}
V(\hat{\bm k},\hat{\bm k}')=-V\sum_\gamma
\varphi_{\gamma}(\hat{\bm k})\varphi^*_{\gamma}(\hat{\bm k}'),
\end{equation}
with $V>0$, where the basis functions satisfy the orthonormal relation, $\left\langle \varphi^*_{\gamma}(\hat{\bm k})\varphi_{\gamma'}(\hat{\bm k}) \right\rangle=\delta_{\gamma\gamma'}$.
Suppose ${\bm\Delta}_{\bm R}(\hat{\bm k})={\bm\Delta}_{\bm R}\varphi_{\gamma}(\hat{\bm k})$, the gap equation is rewritten as
\begin{equation}
V^{-1} {\bm\Delta}_{\bm R} =
\pi T N_0\sum_n\left\langle
\varphi_{\gamma}^*(\hat{\bm k}){\bm f}_{\bm R}(\hat{\bm k};i\omega_n)
\right\rangle.
\label{gap-eq2}
\end{equation}

Equations (\ref{ampere-eq}), (\ref{qc-eq2}) and (\ref{gap-eq2}) constitute the quasiclassical formalism to describe superconducting states under magnetic fields.
Alternatively the system of equations can be derived as the saddle point of the following generating functional per unit volume (measured from the energy in the normal state) \cite{Eilenberger68},
\begin{multline}
\Omega_{\rm SN}\left[
{\bm\Delta}_{\bm R}, {\bm\Delta}^*_{\bm R}, {\bm f}_{\bm R}, {\bm f}_{\bm R}^\dagger, {\bm A}_{\bm R}
\right]=
\\
\int d{\bm R}
\left[
\frac{\left({\bm B}_{\bm R}-{\bm H}_{\bm R}\right)^2}{8\pi}
+V^{-1} \left|{\bm\Delta}_{\bm R}\right|^2
\right.\\ \left.
-2\pi T N_0 \sum_{n\ge 0}
\biggl(
\left\langle I_{\bm R}(\hat{\bm k};i\omega_n)\right\rangle+I'_{\bm R}(i\omega_n)
\biggr)
\right],
\label{free-energy}
\end{multline}
where
\begin{multline}
I_{\bm R}(\hat{\bm k};i\omega_n)=
\varphi^*_{\gamma}(\hat{\bm k}){\bm\Delta}^*_{\bm R}\cdot{\bm f}_{\bm R}
+{\bm f}_{\bm R}^\dagger\cdot{\bm\Delta}_{\bm R}\varphi_{\gamma}(\hat{\bm k})
\\
-\frac{1}{1+g_{\bm R}}\left[
{\bm f}_{\bm R}^\dagger\cdot{\cal L}_+{\bm f}_{\bm R}
+{\bm f}_{\bm R}\cdot{\cal L}_-{\bm f}_{\bm R}^\dagger
\right]
\label{func-i}
\end{multline}
and
\begin{equation}
I'_{\bm R}(i\omega_n)=\frac{1}{2\tau_0\sin^2\delta}\ln[\cos^2\delta+D_{\bm R}\sin^2\delta].
\end{equation}

In eq.~(\ref{free-energy}) the summation over the Matsubara frequencies must be cut off at very large but a finite frequency.
This cut-off procedure can be avoided by a prescription,
\begin{equation}
V^{-1}\to N_0\left[\ln\left(\frac{T}{T_c}\right)+2\pi T\sum_{n=0}^\infty\frac{1}{\omega_n}\right],
\label{cut-off-1}
\end{equation}
namely, the cut-off is absorbed in $T_c$ and the second term cancels the contribution from the third term in eq.~(\ref{free-energy}) at $n\to\infty$.
Note that $T_c$ is the transition temperature at ${\bm H}=0$ without the impurity scattering.

If the quasiclassical equations, eq.~(\ref{qc-eq2}) are solved separately for given ${\bm\Delta}_{\bm R}$, ${\bm\Delta}^*_{\bm R}$ and ${\bm A}_{\bm R}$, eq.~(\ref{func-i}) becomes
\begin{multline}
I_{\bm R}=
{\bm\Delta}^*_{\bm R}(\hat{\bm k})\cdot{\bm f}_{\bm R}
+{\bm f}_{\bm R}^\dagger\cdot{\bm\Delta}_{\bm R}(\hat{\bm k})
+2\sigma^{(n)}_{\bm R}(1-g_{\bm R})
\\
-\frac{g_{\bm R}}{1+g_{\bm R}}\left[
\tilde{\bm\Delta}_{\bm R}^\dagger(\hat{\bm k};i\omega_n)\cdot{\bm f}_{\bm R}+{\bm f}_{\bm R}^\dagger\cdot\tilde{\bm\Delta}_{\bm R}(\hat{\bm k};i\omega_n)
\right]
,
\label{func-i2}
\end{multline}
and $\Omega_{\rm SN}$ coincides with the physical free-energy functional of ${\bm\Delta}_{\bm R}$, ${\bm\Delta}^*_{\bm R}$ and ${\bm A}_{\bm R}$.
In the next section, we show that a simple approximation provides an approximate analytic solution of eq.~(\ref{qc-eq2}).

\section{Approximate analytic solution}
In the previous section, we have derived the set of the quasiclassical equations as the saddle point of the free energy functional.
Basically, a minimum of the free energy functional provides a complete description for thermodynamic properties of superconducting states under magnetic fields.
In this section, however, we show that a simple but a reliable approximation drastically simplifies the problem.
In the following subsection, we show that a kind of ``mean-field" approach leads to an approximate analytic solution for quasiclassical equations (\ref{qc-eq2}).
Then we discuss self-consistent equations for the impurity scattering self-energy.
Lastly, expressions of various thermodynamic quantities are given.

\subsection{BPT approximation}
The approximation used here was first proposed in the Gor'kov formalism by Brandt, Pesch and Tewordt (BPT) \cite{Brandt67}, and later Pesch obtained analytic solutions in the quasiclassical formalism \cite{Pesch75}.
Although BPT initially aimed to describe superconducting states near the upper critical field $H_{c2}$, it is meaningful to extrapolate the method to lower fields.
In this direction, Dahm and co-workers have made a comparison with the numerical solution of quasiclassical equations \cite{Dahm02}.
However, in numerical calculation they assumed the structure of the vortex cores and did not solve whole equations self-consistently, hence a reasonable comparison of the BPT approximation with full numerical solutions is still missing.
After the approximation and the formulation are introduced, we will discuss this point.

In the BPT theory the internal field ${\bm B}_{\bm R}$ is approximated by its spatial average ${\bm B}$.
Hereafter a spatial average of a quantity $X_{\bm R}$ is denoted by $X$ or $\overline{X_{\bm R}}$.
An Abrikosov solution\cite{Tinkham75} is then used for vortex lattice structures for all field range above the lower critical field.
When the internal magnetic field ${\bm B}$ sets parallel to $z$ axis, we write the Abrikosov solution with the Landau gauge ${\bm A}_{\bm R}=(0,Bx,0)$ as
\begin{equation}
\Delta_{\bm R}=\sum_n C_n e^{-ip_ny'}\Phi_0\left[(x'-\Lambda^2p_n)/\Lambda\right],
\label{abrikosov}
\end{equation}
where $p_n=2\pi n/\beta$, $\beta$ being the lattice constant in the $y$ direction and $\Lambda=(2|e|B)^{-1/2}$ is of the order of the lattice spacing of the vortex lattice.
We have taken into account anisotropy of the Fermi velocity in the $x$-$y$ plane by transferring the coordinates as
\begin{equation}
x=\frac{\xi_x}{\xi_{\rm eff}}x',\;\;\;\;\;y=\frac{\xi_y}{\xi_{\rm eff}}y',
\end{equation}
where $\xi_i$ ($i=x,y$) is the $i$-component of the coherence length proportional to the square-root of $\langle v^2_i(\hat{\bm k})\rangle$ and $\xi_{\rm eff}=\sqrt{\xi_x\xi_y}$ sets equal to $\Lambda$.
Here $\Phi_0[x]=\exp(-x^2/2)$ is the lowest eigenfunction of a harmonic oscillator and the periodicity of the coefficients $C_n$ specifies the type of vortex lattice.
Since $\hat{\bm k}$ dependence of the gap function is irrelevant in the following discussion, we omit $\varphi_{\gamma}(\hat{\bm k})$.
For the vector gap, the argument below can be applied separately to each component.

In addition to the above approximations, we expect that the uniform component $g=\overline{g_{\bm R}}$ could describe a global suppression of the gap amplitude as an average.
This is because the higher Fourier ${\bm K}$ components of $g_{\bm R}$ decrease rapidly as $\exp(-\Lambda^2K^2)$ and the normal propagator $g_{\bm R}$ does not contain an important phase variation due to vortices \cite{Brandt67}.
On the other hand, the exact spatial form of ${\bm\Delta}_{\bm R}$ due to its phase winding around vortices will be taken into account in determining the anomalous propagator ${\bm f}_{\bm R}$ \cite{Pesch75}.

Within the above approximation, we can solve analytically the quasiclassical equations (\ref{qc-eq2}).
It is sufficient to solve the first two equations in (\ref{qc-eq2}).
The argument here essentially follows the discussion by Houghton and Vekhter for $s$-wave superconductors with the operator formalism \cite{Houghton98}.
Within the above approximations the anomalous propagator is given by
\begin{equation}
{\bm f}_{\bm R}=g\varphi_{\gamma}(\hat{\bm k})\eta^{(a)}(i\omega_n){\cal \tilde{L}}_+^{-1}{\bm\Delta}_{\bm R}.
\end{equation}
Here we have assumed that the renormalized gap function has the same ${\bm R}$ dependence as the bare gap ${\bm \Delta}_{\bm R}$ and the renormalization factors of the frequency and the gap are independent of ${\bm R}$, i.e.,
\begin{equation}
\tilde{\omega}_n=\eta^{(n)}(i\omega_n)\omega_n,\;\;\;
\tilde{\bm \Delta}_{\bm R}=\eta^{(a)}(i\omega_n){\bm \Delta}_{\bm R}.
\end{equation}
To calculate the inverse of ${\cal \tilde{L}}_+$, we use an operator identity for $\omega_n>0$,
\begin{equation}
{\cal \tilde{L}}_+^{-1}=\int_0^\infty dt \exp[-{\cal \tilde{L}}_+t].
\end{equation}
Then we need to know how the exponential operator acts on the gap.

Since the Abrikosov lattice is a superposition of $\Phi_0$ at different vortex cores, we introduce the lowering and raising operators,
\begin{eqnarray}
&&a=\frac{\Lambda}{\sqrt{2}}\left(\pi_{x'}-i\pi_{y'}\right),
\\&&a^\dagger=\frac{\Lambda}{\sqrt{2}}\left(\pi_{x'}+i\pi_{y'}\right),
\end{eqnarray}
where $\pi'_{i'}=-i\nabla_{i'}-2eA_{i'}=\xi_i\pi_i/\Lambda$ is the dynamical (gauge-invariant) momentum for the Cooper pair.
The operators satisfy the usual bosonic commutation relation $[a,a^\dagger]=1$.
Since $a \Delta_{\bm R}=0$, the Abrikosov lattice (\ref{abrikosov}) can be regarded as the vacuum state $|0\rangle$ in the language of the operator formalism.
Applying $(a^\dagger)^m$ to the vacuum state we obtain the excited states,
\begin{equation}
|m\rangle=\sum_n C_n e^{-ip_ny'}\Phi_m[(x'-\Lambda^2p_n)/\Lambda],
\end{equation}
which corresponds to the $m$-th excited oscillator states centered on each vortices.
Similarly, we introduce the raising and lowering operators as $b^\dagger=(a^\dagger)^*$ and $b=(a)^*$ acting on the conjugate vacuum state, $\Delta_{\bm R}^*=\langle 0|$, since $(a)^*\Delta_{\bm R}^*=0$ holds.
Owing to the factor $e^{ip_ny'}$, a spatial average ensures an orthogonal relation for different excited states as
\begin{equation}
\overline{\langle m | m' \rangle} = |\Delta|^2 \delta_{m,m'},
\end{equation}
since only functions $\Phi_m$ and $\Phi_{m'}$ centered on the same vortex contribute to the integral.
Here we have defined the averaged gap magnitude as $|\Delta|^2=\overline{|\Delta_{\bm R}|^2}$.
Without loss of generality, we can take $\Delta$ real, because $\Delta$ appears only in the form $|\Delta|^2$.
Note that the orthogonality does not hold without the spatial average.

In the operator formalism denoting
\begin{equation}
{\bm v}(\hat{\bm k})=\left[v_\perp(\hat{\bm k})\cos\phi_v,v_\perp(\hat{\bm k})\sin\phi_v,v_\parallel(\hat{\bm k})\right],
\end{equation}
we write
\begin{equation}
{\cal \tilde{L}}_+=\tilde{\omega}_n+\frac{i}{\sqrt{2}}\left(\frac{\tilde{v}_\perp(\hat{\bm k})}{2\Lambda}\right)\left(
e^{-i\phi'}a^\dagger+e^{i\phi'}a\right),
\end{equation}
where we have defined
\begin{eqnarray}
\tilde{v}_\perp(\hat{\bm k})&=&v_\perp(\hat{\bm k})\sqrt{\chi^{-1/2}\cos^2\phi_v+\chi^{1/2}\sin^2\phi_v}\nonumber \\
&=&\sqrt{\chi^{-1/2}v_x^2+\chi^{1/2}v_y^2},\\
\tan\phi'&=&\chi^{1/2}\tan\phi_v=\chi^{1/2}v_y/v_x,
\end{eqnarray}
with $\chi=\langle v_x^2 \rangle / \langle v_y^2 \rangle$.
Note that $\tilde{v}_\perp(\hat{\bm k})$ is of the order of $[\langle v_x^2 \rangle\langle v_y^2 \rangle]^{1/4}$, and we obtain $\tilde{v}_\perp(\hat{\bm k})=v_\perp(\hat{\bm k})$ and $\phi'=\phi_v$ in the absence of the anisotropy, i.e., $\chi=1$.
Using Taylor expansions in $a^\dagger$ and $a$, we obtain
\begin{multline}
{\cal \tilde{L}}_+^{-1}\Delta_{\bm R}=\sqrt{\pi}\left(\frac{2\Lambda}{\tilde{v}_\perp(\hat{\bm k})}\right)
\sum_{m=0}^\infty e^{-im\phi'}
\\ \times
\frac{1}{\sqrt{m!}}\left(-\frac{1}{\sqrt{2}}\right)^mW^{(m)}(i\tilde{u}_n)|m\rangle,
\label{l-inv}
\end{multline}
with $\tilde{u}_n(\hat{\bm k};i\omega_n)=[2\Lambda/\tilde{v}_\perp(\hat{\bm k})]\tilde{\omega}_n$, where $W(z)=e^{-z^2}{\rm erfc}(-iz)$ is the Faddeeva function and $W^{(m)}(z)$ denotes its $m$-th derivative.
The conjugate version ${\cal \tilde{L}}_-^{-1}\Delta_{\bm R}^*$ is obtained by the replacements $\phi'\to-\phi'$ and $|m\rangle\to\langle m|$ in eq.~(\ref{l-inv}).
Then we obtain
\begin{equation}
\overline{{\bm f}_{\bm R}^\dagger\cdot{\bm f}_{\bm R}}=\frac{\sqrt{\pi}}{i}\left(\frac{2\Lambda}{\tilde{v}_\perp(\hat{\bm k})}\right)^2\eta^{(a)\dagger}\eta^{(a)}|{\bm\Delta}(\hat{\bm k})|^2W^{(1)}(i\tilde{u}_n)g^2,
\end{equation}
where we have used the formula,
\begin{equation}
\sum_{m=0}^\infty \frac{1}{m!}\left(\frac{1}{2}\right)^m\left[W^{(m)}(i\tilde{u}_n)\right]^2=
\frac{1}{i\sqrt{\pi}}W^{(1)}(i\tilde{u}_n),
\end{equation}
and we have abbreviated ${\bm\Delta}\varphi_{\gamma}(\hat{\bm k})$ to ${\bm\Delta}(\hat{\bm k})$.

Using the above result together with the normalization, $g^2+\overline{{\bm f}_{\bm R}^\dagger\cdot{\bm f}_{\bm R}}=1$, we finally obtain the normal propagator,
\begin{equation}
g=\left[
1+\frac{\sqrt{\pi}}{i}\left(\frac{2\Lambda}{\tilde{v}_\perp(\hat{\bm k})}\right)^2\eta^{(a)\dagger}\eta^{(a)}|{\bm\Delta}(\hat{\bm k})|^2W^{(1)}(i\tilde{u}_n)
\right]^{-\frac{1}{2}}.
\label{func-p}
\end{equation}
The quantity $I=\overline{I_{\bm R}}$ is also obtained as
\begin{multline}
I=\sqrt{\pi}\left(\frac{2\Lambda}{\tilde{v}_\perp(\hat{\bm k})}\right)|{\bm\Delta}(\hat{\bm k})|^2W(i\tilde{u}_n)g
\biggl[
\eta^{(a)}+\eta^{(a)\dagger}
\\
-\frac{2g}{1+g}\eta^{(a)\dagger}\eta^{(a)}
\biggr]
+2\omega_n(\eta^{(n)}-1)(1-g).
\label{val-i}
\end{multline}
In the clean limit $\nu=1/2\tau_0T_c\ll1$, we have
\begin{equation}
I=\frac{2g}{1+g}\sqrt{\pi}\left(\frac{2\Lambda}{\tilde{v}_\perp(\hat{\bm k})}\right)|{\bm\Delta}(\hat{\bm k})|^2W(iu_n).
\end{equation}
In the absence of $B$ using $W(z)\sim i/\sqrt{\pi}z$ and $W^{(1)}(z)\sim -i/\sqrt{\pi}z^2$ for $|z|\gg1$, we recover the uniform solution of eqs.~(\ref{propa-g0}) and (\ref{propa-f0}).
On the other hand, we get $g=1$ and $I=I'=0$ in the normal state, ${\bm\Delta}=0$.

\subsection{Self-consistent equations for renormalization factors, $\eta^{(n)}(i\omega_n)$ and $\eta^{(a)}(i\omega_n)$}
To complete the system of equations, we need self-consistencies for $\eta^{(n)}(i\omega_n)$ and $\eta^{(a)}(i\omega_n)$.
In the case of non-$s$-wave pairings, we neglect the renormalization of the gap, i.e. $\langle f_{\bm R}\rangle\sim 0$, because the angular average with the factor $\varphi_{\gamma}(\hat{\bm k})$ gives small contribution to the self-energy (the angular average completely vanishes in the absence of magnetic fields).
We then obtain
\begin{equation}
\eta^{(n)}=1+\frac{1}{2\omega_n\tau_0}\frac{\langle g \rangle}{\cos^2\delta+\langle g \rangle^2\sin^2\delta},\;\;\;\eta^{(a)}=1.
\label{renorm-non-s}
\end{equation}
On the other hand, in the case of the isotropic $s$-wave singlet $\varphi_{\gamma}(\hat{\bm k})=1$, we consider only the Born limit $\delta=0$.
Retaining the term proportional to $\Delta_{\bm R}$ ($m=0$ component) in eq.~(\ref{l-inv}), namely,
\begin{equation}
f_{\bm R}\sim g\sqrt{\pi}\left(\frac{2\Lambda}{\tilde{v}_\perp(\hat{\bm k})}\right)W(i\tilde{u}_n)\Delta_{\bm R},
\end{equation}
(this is exact for an isotropic Fermi surface), we get
\begin{eqnarray}
&&
\eta^{(n)}=1+\frac{1}{2\omega_n\tau_0}\langle g \rangle,
\nonumber \\&&
\eta^{(a)}=\left[
1-\frac{\sqrt{\pi}}{2\tau_0}\left\langle g\left(\frac{2\Lambda}{\tilde{v}_\perp(\hat{\bm k})}\right)W(i\tilde{u}_n)\right\rangle\right]^{-1}.
\nonumber \\&&
\label{renorm-s}
\end{eqnarray}
Note that the Anderson theorem ($\eta^{(n)}/\eta^{(a)}=1$) is restored in the case of $B=0$.
Thus in all cases in the limit $B=0$, we recover all the results of the standard theory for dirty superconductors.
In the clean limit we have $\eta^{(n)}(i\omega_n)=\eta^{(a)}(i\omega_n)=1$.

\subsection{Thermodynamic quantities}
With the approximate analytic solutions obtained in the previous subsection, the free energy (\ref{free-energy}) can be regarded as a function of ${\bm\Delta}$ (can be taken as real) and $B$ for given temperature $T$ and the external field $H$.
To obtain an equilibrium state, we minimize the free energy (\ref{free-energy}) with respect to $(B,{\bm \Delta})$.
In searching the minimum point, we need to compute the free energy for given $(B,{\bm \Delta})$.
To do this we first solve the self-consistent equations for the impurity renormalization $\eta^{(n)}(i\omega_n)$ and $\eta^{(a)}(i\omega_n)$ at positive $\omega_n$'s, i.e. eqs.~(\ref{func-p}) and (\ref{renorm-non-s}) for non-$s$-wave pairings, or eqs.~(\ref{func-p}) and (\ref{renorm-s}) for $s$-wave singlet.
This can easily be done by an iteration starting from the clean limit, $\eta^{(n)}=\eta^{(a)}=1$.
Once we determine $\eta^{(n)}$ and $\eta^{(a)}$, we can compute the free energy (\ref{free-energy}) via eqs.~(\ref{func-p}) and (\ref{val-i}).
Note that all the numerical calculations can be done with real quantities in the Matsubara formalism, since $i^nW^{(n)}(ix)$ is real for real $x$.

Once we obtain the equilibrium values of $B$ and ${\bm\Delta}$, we can compute various thermodynamic quantities as the spatial average.
The magnetization is given by
\begin{equation}
-4\pi M(T,H)=H-B(T,H).
\end{equation}
Viewing the one-body nature of the quasiparticles in the present formalism, we can write down the entropy,
\begin{multline}
S(T,H)=-4\int_0^\infty d\omega N(\omega;T,H)
\biggl[
f\left(\frac{\omega}{T}\right) \ln f\left(\frac{\omega}{T}\right)
\\
+ \left(1-f\left(\frac{\omega}{T}\right)\right)\ln \left(1-f\left(\frac{\omega}{T}\right)\right)
\biggr],
\label{entropy}
\end{multline}
with $f(x)=(1+e^x)^{-1}$ via the DOS in the superconducting state,
\begin{equation}
\frac{N(\omega;T,H)}{N_0}={\rm Re}\left\langle g(\hat{\bm k};i\omega_n\to\omega+i\delta) \right\rangle,
\end{equation}
where we have done the analytic continuation from $i\omega_n$ in the upper half plane to $\omega+i\delta$, $\delta$ being an infinitesimal positive number.
Then the specific heat is calculated numerically by $C(T,H)/T=\partial S/\partial T$.

It is useful to discuss the region $B\sim H\sim H_{c2}$ in the clean limit, where ${\bm \Delta}$ is suppressed.
Expanding the free energy in ${\bm\Delta}$, we obtain the free energy under magnetic fields,
\begin{equation}
\Omega_{\rm SN}\sim\frac{(B-H)^2}{8\pi}+N_0\alpha(T,B)|{\bm\Delta}|^2+\frac{1}{2}\frac{N_0}{T_c^2}\beta(T,B)\left(|{\bm\Delta}|^2\right)^2,
\end{equation}
where
\begin{multline}
\alpha(T,B)=\biggl\langle
\biggl[
\ln\left(\frac{T}{T_c}\right)+2\pi T
\sum_{n=0}^\infty
\biggl\{
\frac{1}{\omega_n}
\\
-\sqrt{\pi}\left(\frac{2\Lambda}{\tilde{v}_\perp(\hat{\bm k})}\right)W(iu_n)\biggr\}\biggr]|\varphi_{\gamma}(\hat{\bm k})|^2
\biggr\rangle,
\label{alpha}
\end{multline}
and
\begin{multline}
\beta(T,B)=-i\pi^2TT_c^2\sum_{n=0}^\infty
\left\langle \left(\frac{2\Lambda}{\tilde{v}_\perp(\hat{\bm k})}\right)^3
\right.\\ \times\left.
W(iu_n)W^{(1)}(iu_n)|\varphi_{\gamma}(\hat{\bm k})|^4\right\rangle.
\label{beta}
\end{multline}
The temperature dependence of the upper critical field is determined by
\begin{equation}
\alpha(T,H_{c2})=0.
\end{equation}
In the limit $H_{c2}\to0$ we recover the BCS results, i.e. $\alpha=\ln(T/T_c)$ and $\beta=[7\zeta(3)/8\pi^2]\langle |\varphi_{\gamma}(\hat{\bm k})|^4\rangle$, where $\zeta(n)$ is the Riemann zeta function.
The field dependence of the magnetization and the gap near $H_{c2}$ are given as
\begin{multline}
-4\pi M=4\pi\alpha_{c2}'(T)N_0|{\bm\Delta}|^2=
\\
\left(\frac{\beta_{c2}(T)}{4\pi\alpha_{c2}'^2(T)N_0T_c^2}-1\right)^{-1}(H_{c2}-H),
\label{magnetization}
\end{multline}
where $\alpha_{c2}'(T)=\partial\alpha(T,B)/\partial B|_{B=H_{c2}}$ and $\beta_{c2}(T)=\beta(T,H_{c2})$.
At $T=0$ we can perform the summation over $\omega_n$ in (\ref{alpha}) and (\ref{beta}).
Then we obtain
\begin{equation}
\alpha(0,B)=\frac{1}{2}\ln\left(\frac{B}{H_{c2}}\right),
\end{equation}
with the upper critical field at $T=0$,
\begin{equation}
H_{c2}=\frac{2}{|e|}\left(\frac{\pi v}{T_c}\right)^2\exp\left[-2\left\langle|\varphi_{\gamma}(\hat{\bm k})|^2\ln\left(\frac{\tilde{v}_\perp(\hat{\bm k})}{v}\right)\right\rangle-\gamma\right],
\end{equation}
where $v$ without $(\hat{\bm k})$ denotes the angular average of $|{\bm v}(\hat{\bm k})|$ and $\gamma\doteqdot0.577216$ is the Euler's constant, and
\begin{equation}
\beta(0,B)=\frac{\pi}{2|e|B}\left(\frac{T_c}{v}\right)^2
\left\langle
\left(\frac{v}{\tilde{v}_\perp(\hat{\bf k})}\right)^2
|\varphi_\gamma(\hat{\bf k})|^4
\right\rangle.
\end{equation}
In the limit $\chi\ll1$ using $\tilde{v}_\perp/v_\perp\sim\chi^{1/4}$, we obtain
\begin{equation}
\frac{H_{c2}(T=0;\chi)}{H_{c2}(T=0;1)}\sim \chi^{-1/2}=\sqrt{\frac{\langle v_y^2 \rangle}{\langle v_x^2 \rangle}}.
\end{equation}

\section{Extension to multiband superconductors}
In a system having plural Fermi surfaces, each band could have different symmetry and/or magnitude of gaps due to symmetry reason \cite{Suhl59,Kondo63,Leggett66}.
Since those Fermi surfaces have different shape in general, the Cooper pair may be formed among the same band in order to reduce the energy associated with the center-of-mass motion.
In this case the MF Hamiltonian (\ref{mf-ham}) is diagonal in the band indices $\ell$, and the quasiclassical propagators are also diagonal.
Moreover, we do not take into account the interband impurity scattering.
It is known that the mixing effect between different bands disappears both in the weak and the strong-coupling limit of the interband impurity scattering, hence the multiband situation is restored \cite{Kulic99,Ohashi02}.
Therefore, the coupling between bands only appears in the gap equation as
\begin{equation}
\sum_{\ell'}\left(V^{-1}\right)_{\ell\ell'} {\bm\Delta}_{\ell'{\bm R}} =
\pi T N_{0\ell}\sum_n\left\langle
\varphi_{\ell\gamma}^*(\hat{\bm k}){\bm f}_{\ell{\bm R}}(\hat{\bm k};i\omega_n)
\right\rangle,
\end{equation}
with ${\bm\Delta}_{\ell{\bm R}}(\hat{\bm k})={\bm\Delta}_{\ell{\bm R}}\varphi_{\ell\gamma}(\hat{\bm k})$, which corresponds to eq.~(\ref{gap-eq2}).
Here we have used
\begin{equation}
V_{\ell\ell'}(\hat{\bm k},\hat{\bm k}')=-\left(V \right)_{\ell\ell'}\sum_\gamma
\varphi_{\ell\gamma}(\hat{\bm k})\varphi^*_{\ell'\gamma}(\hat{\bm k}'),
\end{equation}
where the coupling constant $V$ is a matrix in the band space.

Now we express a set of formulas in terms of dimensionless parameters.
The free energy scaled by $N_0T_c^2$, $N_0=\sum_\ell N_{0\ell}$ being the total DOS in the normal state, is given by
\begin{multline}
\frac{\Omega_{\rm SN}}{N_0T_c^2}=
\frac{\kappa^2}{8\pi}(b-h)^2+\sum_{\ell\ell'}\alpha^{(0)}_{\ell\ell'}{\bm d}^*_\ell\cdot{\bm d}_{\ell'}
\\
+\sum_\ell n_\ell\left[|{\bm d}_\ell|^2\ln t+2\pi t\sum_{n=0}^{\infty}\left(
\frac{|{\bm d}_\ell|^2}{o_n}-\left\langle i_\ell \right\rangle-i'_\ell
\right)
\right],
\end{multline}
with
\begin{equation}
\alpha^{(0)}_{\ell\ell'}=(\lambda^{-1})_{\ell\ell'}-n_\ell A_c\delta_{\ell\ell'},
\end{equation}
where $\lambda=N_0 V$, $n_\ell=N_{0\ell}/N_0$, $(b,h)=(B,H)/H_{c2}(T=0)$ and $(t,{\bm d}_\ell,o_n,i_\ell,i'_\ell)=(T,{\bm\Delta}_\ell,\omega_n,I_\ell,I'_\ell)/T_c(H=0)$.
Since all the argument in the previous section holds for each band $\ell$, we have added the subscript $\ell$ to the relevant quantities.
Note that the angular average should be carried out over the corresponding Fermi surface.
Here the quantity defined as $\kappa=H_{c2}/T_c\sqrt{N_0}$ is the GL parameter, which is of the order of the ratio between the penetration depth and the coherence length of the primary order parameter (specified by $\ell=1$).
The factor $A_c=\ln(2e^\gamma\omega_c/\pi T_c)$ is the smallest eigenvalue determined by
\begin{equation}
{\rm det}\;\alpha^{(0)}_{\ell\ell'}=0.
\end{equation}
It is useful to introduce a dimensionless critical field $b_{c2}=|e|(v/T_c)^2H_{c2}/2$, which will be determined such that the gap vanishes at $b=1$ and $t=0$.
We summarize appropriate formulas for (a) non $s$-wave pairings, (b) $s$-wave singlet, and (c) the clean limit.
\begin{enumerate}
\item[(a)] non $s$-wave pairings,
\begin{equation}
g_\ell=\left[
1+\frac{\sqrt{\pi}}{i}\left(\frac{v_1}{\tilde{v}_{\ell\perp}(\hat{\bm k})}\right)^2\frac{|{\bm d}_\ell(\hat{\bm k})|^2}{bb_{c2}}W^{(1)}(i\tilde{u}_{\ell n})
\right]^{-1/2},
\end{equation}
\begin{equation}
\eta^{(n)}_{\ell}=1+\frac{\nu_\ell}{o_n}\frac{\langle g_\ell \rangle}{\cos^2\delta_\ell+\langle g_\ell \rangle^2\sin^2\delta_\ell},
\end{equation}
and
\begin{multline}
i_\ell=\frac{2g_\ell}{1+g_\ell}\sqrt{\pi}\left(\frac{v_1}{\tilde{v}_{\ell\perp}(\hat{\bm k})}\right)\frac{|{\bm d}_\ell(\hat{\bm k})|^2}{\sqrt{bb_{c2}}}W(i\tilde{u}_{\ell n})
\\
+2o_n(\eta^{(n)}_\ell-1)(1-g_\ell),
\end{multline}
\begin{equation}
i'_\ell=\frac{\nu_\ell}{\sin^2\delta_\ell}\ln\left[\frac{\nu_\ell\langle g_\ell \rangle}{o_n(\eta^{(n)}_\ell-1)}\right],
\end{equation}
\item[(b)] $s$-wave singlet
\begin{equation}
g_\ell=\left[
1+\frac{\sqrt{\pi}}{i}\left(\frac{v_1}{\tilde{v}_{\ell\perp}(\hat{\bm k})}\right)^2\frac{(d_\ell\eta^{(a)}_\ell)^2}{bb_{c2}}W^{(1)}(i\tilde{u}_{\ell n})
\right]^{-1/2},
\end{equation}
\begin{eqnarray}
&&
\eta^{(n)}_{\ell}=1+\frac{\nu_\ell}{o_n}\langle g_\ell \rangle,
\\&&
\eta^{(a)}_{\ell}=\left[ 1-\frac{\sqrt{\pi}\nu_\ell}{\sqrt{bb_{c2}}}\left\langle g_\ell \left(\frac{v_1}{\tilde{v}_{\ell\perp}(\hat{\bm k})}\right)W(i\tilde{u}_{\ell n}) \right\rangle\right]^{-1},
\nonumber\\&&
\end{eqnarray}
and
\begin{multline}
i_\ell=2\sqrt{\pi}\left(\frac{v_1}{\tilde{v}_{\ell\perp}(\hat{\bm k})}\right)\frac{|d_\ell|^2}{\sqrt{bb_{c2}}}W(i\tilde{u}_{\ell n})\left[1-\frac{g_\ell\eta^{(a)}_\ell}{1+g_\ell}\right]\eta^{(a)}_\ell
\\
+2o_n(\eta^{(n)}_\ell-1)(1-g_\ell),
\end{multline}
\begin{equation}
i'_\ell=\frac{1}{\nu_\ell}\left[
o_n^2(\eta^{(n)}_\ell-1)^2+d_\ell^2(1-1/\eta^{(a)}_\ell)^2
\right],
\end{equation}
\item[(c)] clean limit
\begin{equation}
g_\ell=\left[
1+\frac{\sqrt{\pi}}{i}\left(\frac{v_1}{\tilde{v}_{\ell\perp}(\hat{\bm k})}\right)^2\frac{|{\bm d}_\ell(\hat{\bm k})|^2}{bb_{c2}}W^{(1)}(iu_{\ell n})
\right]^{-1/2},
\end{equation}
and
\begin{equation}
i_\ell=\frac{2g_\ell}{1+g_\ell}\sqrt{\pi}\left(\frac{v_1}{\tilde{v}_{\ell\perp}(\hat{\bm k})}\right)\frac{|{\bm d}_\ell(\hat{\bm k})|^2}{\sqrt{bb_{c2}}}W(iu_{\ell n})
\end{equation}
\end{enumerate}
where we have defined the impurity scattering rate as $\nu_\ell=1/2\tau_{\ell 0}T_c$ and $\tilde{u}_{\ell n}=[v_1/\tilde{v}_{\ell\perp}(\hat{\bm k})]\tilde{o}_n/\sqrt{bb_{c2}}$.

In order to discuss properties under the rotating magnetic field, we move to the new coordinate $(x',y',z')$ in which the applied magnetic field is parallel to $z'$ axis.
Representing the rotation matrix $\hat{R}$ for the original magnetic field ${\bm H}=H(\sin\theta_h\cos\phi_h,\sin\theta_h\sin\phi_h,\cos\theta_h)$ as,
\begin{equation}
\hat{R}=\left(
\begin{array}{ccc}
\cos\theta_h\cos\phi_h & \cos\theta_h\sin\phi_h & -\sin\theta_h \\
-\sin\phi_h & \cos\phi_h & 0 \\
\sin\theta_h\cos\phi_h & \sin\theta_h\sin\phi_h & \cos\theta_h
\end{array}
\right),
\end{equation}
we obtain the velocity in the new coordinate as ${\bm v}'_\ell(\hat{\bm k})=\hat{R}{\bm v}_\ell(\hat{\bm k})$.
In the case of ${\bm H}\parallel x$ for example, we have $v_{x'}^2=v_z^2$ and $v_{y'}^2=v_y^2$ with $\theta_h=\pi/2$ and $\phi_h=0$ as they should.
Then, we can express the necessary quantities in terms of ${\bm v}'_\ell(\hat{\bm k})$ as
\begin{eqnarray}
&&\chi_\ell=\langle v_{\ell x'}^2(\hat{\bm k}) \rangle / \langle v_{\ell y'}^2(\hat{\bm k}) \rangle,\\
&&\tilde{v}_{\ell\perp}(\hat{\bm k})=\sqrt{\chi_\ell^{-1/2}v_{\ell x'}^2(\hat{\bm k})+\chi_\ell^{1/2}v_{\ell y'}^2(\hat{\bm k})}.
\end{eqnarray}
For details of ${\bm v}_\ell(\hat{\bm k})$ we can refer to band structure calculations or simple tight-binding models.

Using the argument similar to the single-band case, we obtain the free energy in the clean limit,
\begin{multline}
\frac{\Omega_{\rm SN}}{N_0T_c^2}\sim\frac{\kappa^2}{8\pi}(b-h)^2+\sum_{\ell\ell'}\left(\alpha_{\ell\ell'}^{(0)}+\alpha_\ell(t,b) \delta_{\ell\ell'}\right)
{\bm d}^*_\ell\cdot{\bm d}_{\ell'}
\\
+\frac{1}{2}\sum_\ell \beta_\ell(t,b)\left( |{\bm d}_\ell|^2 \right)^2,
\end{multline}
where
\begin{multline}
\alpha_\ell(t,b)=n_\ell
\biggl\langle
\biggl[
\ln t + 2\pi t \sum_{n=0}^\infty
\biggl\{
\frac{1}{o_n}-\frac{\sqrt{\pi}}{\sqrt{bb_{c2}}}\left(\frac{v_1}{\tilde{v}_{\ell\perp}(\hat{\bm k})}\right)
\\ \times
W(iu_{\ell n})
\biggr\}
\biggr]
|\varphi_{\ell\gamma}(\hat{\bm k})|^2
\biggr\rangle,
\end{multline}
and
\begin{multline}
\beta_\ell(t,b)=-n_\ell i\pi^2t \sum_{n=0}^\infty
\left\langle
(bb_{c2})^{-3/2}\left(\frac{v_1}{\tilde{v}_{\ell\perp}(\hat{\bm k})}\right)^3
\right.\\ \times
W(iu_{\ell n})W^{(1)}(iu_{\ell n})|\varphi_{\ell\gamma}(\hat{\bm k})|^4
\biggr\rangle.
\end{multline}
Note that only the quadratic term contains the coupling between different bands, which corresponds to the internal Josephson coupling.
We obtain the temperature dependence of the upper critical field, $b(t)$ by solving the equation,
\begin{equation}
{\rm det}\left(
\alpha^{(0)}_{\ell\ell'}+\alpha_\ell(t,b) \delta_{\ell\ell'}
\right) = 0.
\end{equation}
In particular, at $T=0$ we have
\begin{multline}
\alpha_{\ell}(0,b)=
\\
n_\ell
\left\langle
\ln
\left[
\frac{e^{\gamma/2}}{\pi}\sqrt{bb_{c2}}
\left(\frac{\tilde{v}_{\ell\perp}(\hat{\bm k})}{v_1}\right)
\right]
|\varphi_{\ell\gamma}(\hat{\bm k})|^2
\right\rangle,
\end{multline}
and
\begin{equation}
\beta_\ell(0,b)=n_\ell\frac{\pi}{4bb_{c2}}
\left\langle
\left(
\frac{v_1}{\tilde{v}_{\ell\perp}(\hat{\bm k})}
\right)^2|\varphi_{\ell\gamma}(\hat{\bm k})|^4
\right\rangle.
\end{equation}

Similarly, we obtain the entropy as
\begin{multline}
\frac{S(t,h)}{N_0T_c}=-4\int_0^\infty dx \frac{N(x;t,h)}{N_0}
\biggl[
f\left(\frac{x}{t}\right) \ln f\left(\frac{x}{t}\right)
\\
+ \left(1-f\left(\frac{x}{t}\right)\right)\ln \left(1-f\left(\frac{x}{t}\right)\right)
\biggr],
\end{multline}
with the total DOS,
\begin{equation}
\frac{N(x;t,h)}{N_0}=\sum_\ell n_\ell {\rm Re}\left\langle
g_\ell(\hat{\bm k};io_n\to x+i\delta)\right\rangle,
\end{equation}
and the specific heat is given by $C(t,h)=t(\partial S/\partial t)$.
For reference, we have $S_{\rm N}(t,h)=C_{\rm N}(t,h)=2\pi^2N_0T/3$ in the normal state.

\section{examples}
Now we demonstrate the application of our theory to typical pairings, i.e. (a) the single-band isotropic $s$-wave, (b) the single-band $d_{x^2-y^2}$-wave and (c) the two-band $s$-waves.
For simplicity, we consider strongly type-II superconductors ($\kappa\gg1$) in the clean limit with two-dimensional cylindrical Fermi surfaces.
The magnetic field is applied along $z$ axis.
We also use the same Fermi velocity for two bands.
Thus we have $B=H$, $\chi=1$, $v_1/\tilde{v}_{\ell\perp}(\hat{\bm k})=1$, $\langle\cdots\rangle=\int_0^{2\pi}\frac{d\phi}{2\pi}\cdots$, and only the gap(s) are to be determined.

\begin{figure*}
\includegraphics[width=14cm]{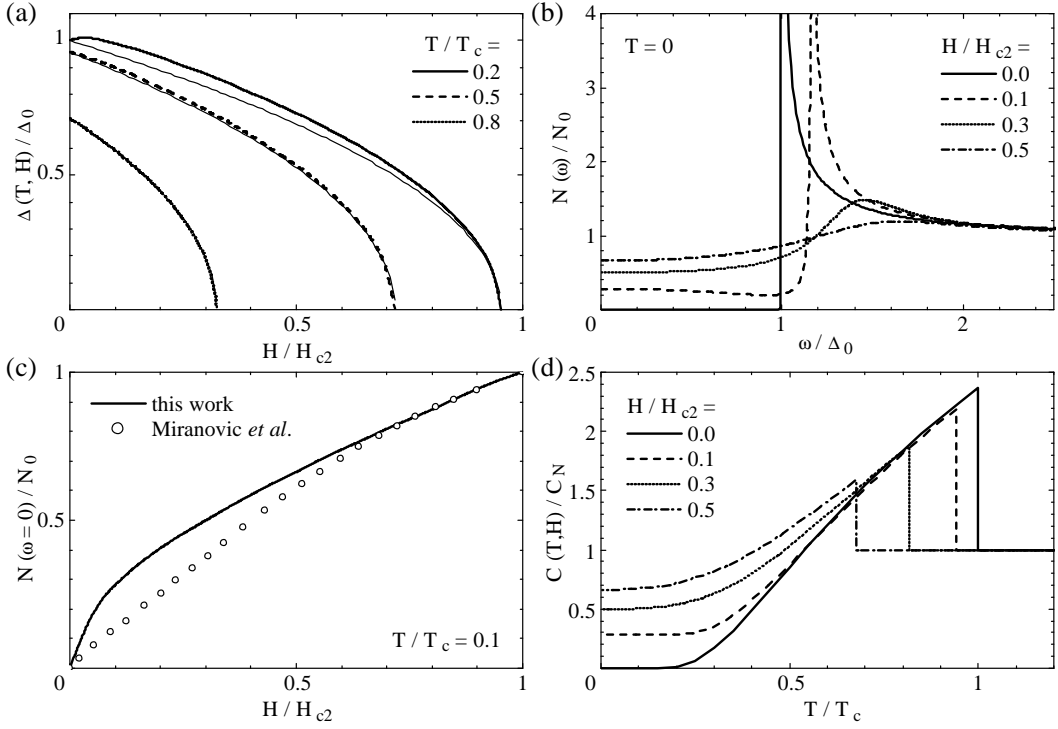}
\caption{The averaged thermodynamic quantities for the $s$-wave: (a) the gap magnitude scaled by $\Delta_0=\Delta(T=H=0)$. The thin line represents the empirical formula $\Delta(T,H)=\Delta(T,0)\sqrt{1-H/H_{c2}}$, (b) the density of states at $T=0$, (c) the field dependence of the zero-energy DOS compared with full numerical results \cite{Miranovic03,Nakai04}, (d) the specific heat.}
\label{fig1}
\end{figure*}

Let us first show results for the single-band isotropic $s$-wave, $\varphi(\hat{\bm k})=1$.
Figure~\ref{fig1} shows the averaged thermodynamic quantities in the mixed state, (a) the field dependence of the gap magnitude $\Delta$ scaled by $\Delta_0=\Delta(h=t=0)$ at different temperatures, (b) the field dependence of the DOS at $t=0$, (c) the field dependence of the zero-energy (ZE) DOS at $t=0.1$, and (d) the temperature dependence of the specific heat normalized to that of the normal state, $C_N$.
In Fig.~\ref{fig1}(a) the gap at the low temperature first increases as the field increases.
This is an drawback of the present approximation, which is not valid at very low fields at low temperature.
The thin line represents an empirical formula, $\Delta(t,h)=\Delta(t,0)\sqrt{1-h}$, which well describes the field dependence of the gap for $t>0.5$.
In Fig.~\ref{fig1}(b) the peak position of the DOS moves upward in energy with the external field.
In Fig.~\ref{fig1}(c) the open circles are taken from the results of the full numerical solution of the quasiclassical equations done by Miranovi\'c {\it et al.} \cite{Miranovic03,Nakai04}.
It indicates that the BPT approximation works very well for $h>0.6$ at low temperatures.
For lower fields the BPT approximation overestimates contribution from vortex cores since the vortex core size becomes unphysically large in the Abrikosov lattice model, (\ref{abrikosov}) (see numerical results \cite{Miranovic03}).

\begin{figure*}
\includegraphics[width=14cm]{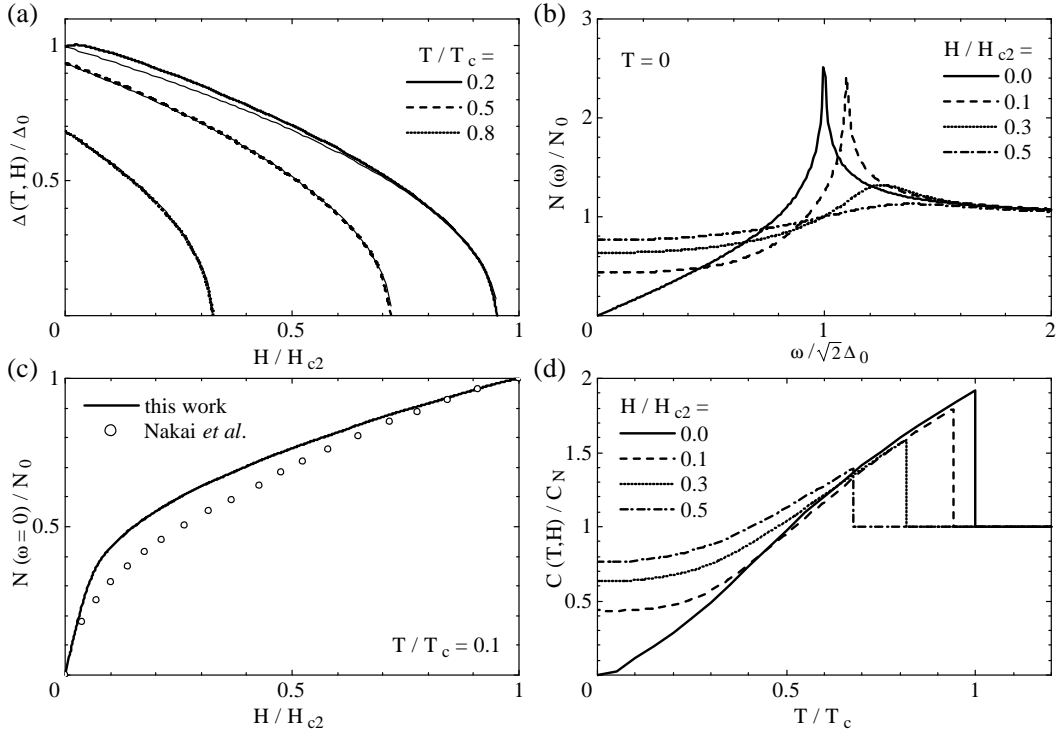}
\caption{The averaged thermodynamic quantities for the $d_{x^2-y^2}$-wave: (a) the gap magnitude. Thin line is the same empirical formula in Fig.~\ref{fig1}, (b) the DOS at $T=0$, (c) the ZEDOS and full numerical results \cite{Nakai04}, (d) the specific heat.}
\label{fig2}
\end{figure*}

Next we move to the case of $d_{x^2-y^2}$-wave, $\varphi(\hat{\bm k})=\sqrt{2}\cos(2\phi)$.
Figure~\ref{fig2} shows results for $d_{x^2-y^2}$-wave in the plot similar to the previous case.
The same empirical formula works fine for $t>0.5$.
A tendency of the peak shift in the DOS is the similar to $s$-wave, but an amount of the shift is smaller than that of $s$-wave.
Although the BPT approximation overestimates the vortex core contribution in comparison with full numerical results \cite{Nakai04,Nakai-c}, it shows better agreement than the case of $s$-wave.

\begin{figure*}
\includegraphics[width=14cm]{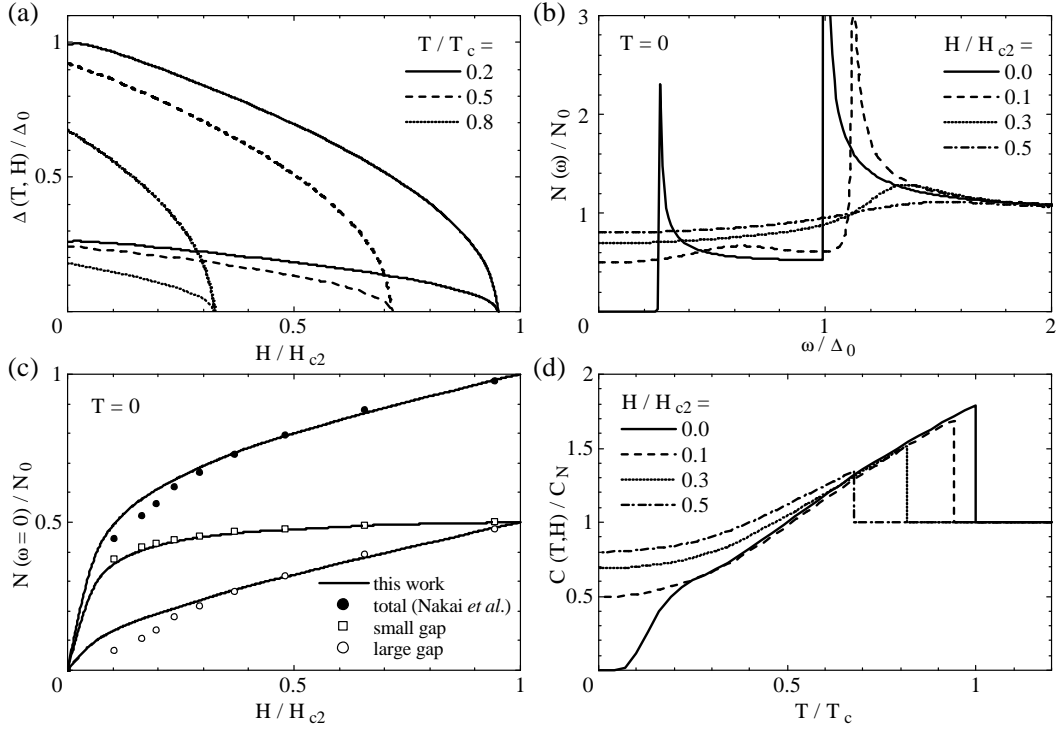}
\caption{The averaged thermodynamic quantities for the two-band $s$-wave: (a) the gap magnitude scaled by the primary gap $\Delta_{0}=\Delta_1(T=H=0)$, (b) the DOS at $T=0$, (c) the ZEDOS and numerical results \cite{Nakai02}, (d) the total specific heat.
The coupling matrix is given by $\hat{\lambda}=((\lambda_1,\lambda),(\lambda,\lambda_2))$ with $\lambda_1=0.25$, $\lambda_2=0.08\lambda_1$ and $\lambda=0.26\lambda_1$.}
\label{fig3}
\end{figure*}

Figure~\ref{fig3} shows results for two-band $s$-waves.
We use the same density of states for both bands, $n_1=n_2=0.5$ for simplicity.
Here the coupling matrix is given by
\begin{equation}
\lambda_{\ell\ell'}=\left(
\begin{array}{cc}
\lambda_1 & \lambda \\ \lambda & \lambda_2
\end{array}
\right),
\end{equation}
with $\lambda_1=0.25$, $\lambda_2=0.08\lambda_1$ and $\lambda=0.26\lambda_1$.
In Fig.~\ref{fig3}(c) the ZEDOS's are compared with numerical calculation of the Bogoliubov de Gennes framework at $T=0$ \cite{Nakai02}.
The field dependence in the passive band agrees very well with the full numerical result since the vortex core size is sufficiently large due to the smallness of the gap with large coherence length.
The discrepancy mainly comes from the primary band but its contribution accounts for $n_1$ of the total ZEDOS.
Therefore, the discrepancy in total becomes less remarkable.
It should be emphasized that behaviors of two-band systems would be changed sensitively by a slight change of material parameters and a combination of two gaps.
Therefore in discussing experimental data it is necessary to take into account precise material parameters in accordance with band-structure calculations \cite{Koshelev03,Dahm03,Zhitomirsky04}.

\begin{figure}
\includegraphics[width=8cm]{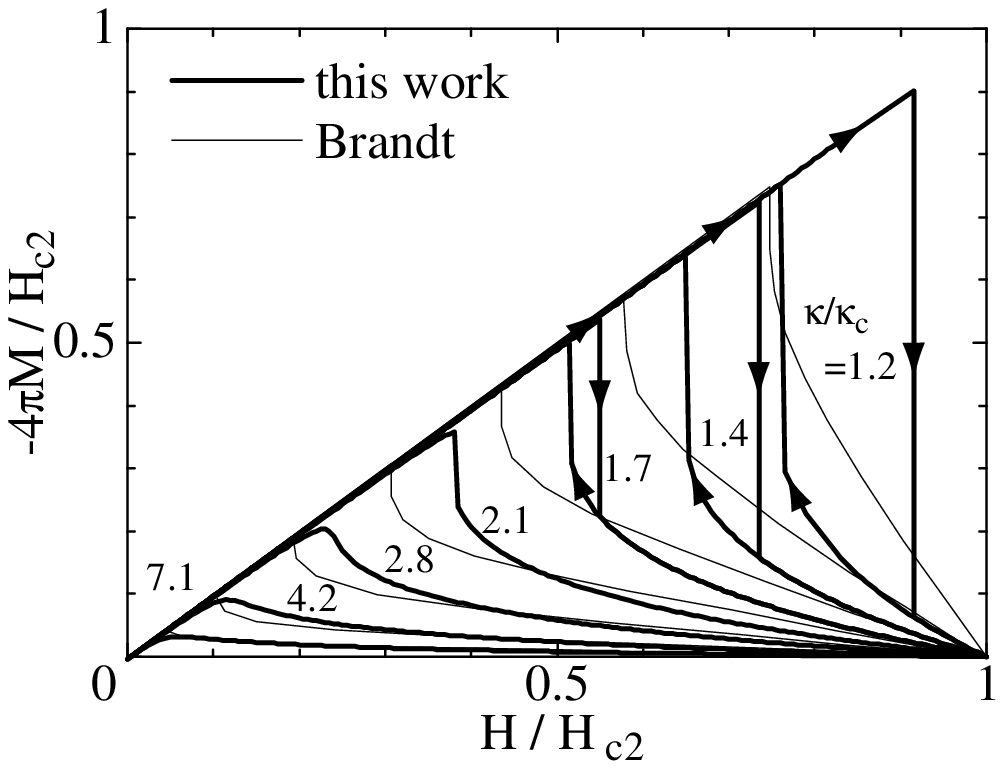}
\caption{The magnetization curves for several GL parameters. The thin lines are taken from the GL result \cite{Brandt03}. For $\kappa/\kappa_c<1.9$ the lower transition becomes first order in the BPT approximation. The arrow indicates the hysteresis loop.}
\label{fig4}
\end{figure}

Finally, we demonstrate the magnetization curves for several $\kappa$'s in the case of $s$-wave in Fig.~\ref{fig4}.
$\kappa_c$ is the critical value of type I-II superconductivity, and $\kappa_c=2\pi^{3/2}e^{-\gamma}\doteqdot6.25278$ in our definition of $\kappa$.
The thin lines denote results obtained by means of the GL theory \cite{Brandt03}.
For larger $\kappa$, the BPT approximation shows better agreement with the GL results.
In the BPT approximation, the transition at the lower critical field, $H_{c1}$ becomes first order for $\kappa/\kappa_c<1.9$, i.e., the free energy has two minima in $(B,\Delta)$ space above $H_{c1}$.
In this case, the hysteresis loop is represented by the arrow.
For small $\kappa$ nonlocal effect becomes important.
Although the BPT approximation properly takes into account nonlocal effect beyond the GL theory, the treatment of averaged magnetic field itself becomes poor.
To conclude the validity of the BPT approximation for small $\kappa$, we should await results of full numerical calculation.

\section{summary}
We have presented a simple calculational scheme of thermodynamic quantities for singlet and unitary triplet states under magnetic fields.
A combination of the approximate analytic solution with a free energy functional in the quasiclassical theory provides a wide use formalism including the impurity scattering and the multiband superconductivity.
We have discussed the simple formula for the upper critical fields in terms of the microscopic free energy under magnetic fields.
The theory requires a little numerical computation as easy as the usual BCS theory without magnetic fields.
We have demonstrated the application to $s$-wave, $d_{x^2-y^2}$-wave and two-band $s$-wave cases.
The comparisons with reliable numerical calculations conclude that the BPT approximation works better for gap with line of nodes than full gap.
It also works well for multiband superconductivity because of its smallness of the gap magnitude in passive band.
The flexibility of the theory allows us to plug in detail structure of the Fermi surface with help of band calculations in an arbitrary direction of the magnetic field.
All the feature of the present theory is appropriate to analyze experimental data semi-quantitatively, especially in the oriented magnetic field.
The method is also useful to discuss linear response and transport coefficients, which will be discussed in a future publication.

\section*{ACKNOWLEDGMENTS}
The author would like to thank I. Vekhter, M. Matsumoto, T. Dahm, D.F. Agterberg, M. Sigrist, T.M. Rice, K. Izawa, N. Nakai and Y. Matsuda for fruitful discussions.
A part of the present work was done in the warm hospitality during his stay in the Institut f\"ur Theoretische Physik, ETH Zurich, Switzerland.
This work was also supported by the NEDO of Japan and Swiss National Fund.

\end{document}